\documentclass[fleqn,10pt]{wlscirep}
\usepackage[utf8]{inputenc}
\usepackage{multirow} 
\usepackage[T1]{fontenc}
\title{M6: Multi-generator, Multi-domain, Multi-lingual and cultural, Multi-genres, Multi-instrument Machine-Generated Music Detection Databases}

\author[1,*,+]{Yupei Li}
\author[2,+]{Hanqian Li}
\author[1]{Lucia Specia}
\author[1,3]{Bj\"orn Schuller}

\affil[1]{Imperial College London, Computing, London, United Kingdom}
\affil[2]{Shandong University, Shandong, China}
\affil[3]{Technical Unitervisty of Munich, Munich, German}

\affil[*]{corresponding author yl7622@ic.ac.uk}

\affil[+]{these authors contributed equally to this work}

\begin{abstract}
Machine-generated music (MGM) has become a powerful tool with applications in music therapy, personalised editing, and creative inspiration. However, its unregulated use threatens the entertainment, education, and arts sectors by diminishing the value of high-quality human compositions. Effective detection of machine-generated music (MGMD) is essential, yet progress is hindered by the lack of comprehensive datasets. To address this gap, we introduce \textbf{M6}, a large-scope benchmark dataset designed for MGMD research. M6 is distinguished by its diversity, encompassing multiple generators, domains, languages, cultural contexts, genres, and instruments, all provided in WAV format. We detail the data collection methodology and analysis, alongside baseline performance scores from foundational binary classification models, highlighting the complexity of MGMD and the need for further advancements. M6 serves as a robust resource to support future research in developing effective detection methods. The dataset is available at \href{https://huggingface.co/datasets/yl7622/M6}{Hugging Face} to promote collaboration and innovation.

\end{abstract}
\begin{document}

\flushbottom
\maketitle
%
%
\thispagestyle{empty}

\section{Introduction}
\label{sec:intro} 
Machine-generated music (MGM) has rapidly gained traction through platforms such as OpenAI’s MuseNet \cite{openai_musenet}, and Jukedeck \cite{jukedeck_tiktok}, driven by advancements in large language models (LLMs), particularly acoustic LLMs. This technology presents huge opportunities and challenges for the music industry, raising concerns about originality, copyright, and artistic value \cite{henry2024impacts}. While MGM can enhance music production by inspiring creativity, providing structural frameworks, and suggesting instrumental arrangements \cite{hadjeres2017deepbach}, its rapid and scalable output risks undermining traditional, artist-created works that rely on time, effort, and personal experience. Moreover, the prevalence of AI-generated styles may unintentionally influence artists, reducing originality, and shift public taste by saturating the music landscape with AI-driven patterns. Consequently, effective MGM detection (MGMD) is essential to safeguarding the creative integrity of the music community and ensuring a balanced coexistence of human and MGM.

Recent works \cite{li2024audio} reviewed the current research of MGMD, together with other music detection tasks \cite{shirol2023comprehensive,dhanapala2024music,dash2023comprehensive} emphasising the urgent need for further development in this area.
The authors highlight critical gaps, noting that the only publicly available dataset, FakeMusicCaps \cite{comanducci2024fakemusiccaps}, was not designed for MGMD tasks, but for generator identification. Another dataset, SONICS \cite{rahman2024sonics}, has its own dataset generation strategy does not sufficiently cover the diversity of music. Given that music is inherently subjective and an art form with diverse formats, providing a sufficiently varied dataset is essential for training effective MGMD models. Such a dataset would facilitate the advancement of MGMD.

Additionally, existing literature suggests that MGMD is substantially more complex than a straightforward binary classification task \cite{cooke2024good}. As noted by Li et al. \cite{li2024audio}, it is rare for contemporary music to be produced without the assistance of computers in arranging and mixing, transforming melody notes into fully realised, performance-ready music. This technological integration can significantly expedite the music production process \cite{martinez2022automatic} and allows for quick adaptation and transformation into desired styles \cite{chi2025music}. Furthermore, art requires inspiration, which has traditionally been derived from other art forms or life experiences. However, AI-generated suggestions can now provide inspiration almost instantaneously \cite{haase2023art,barale2021inspires}. For instance, creators can input specific requirements such as theme, style, or instruments into an AI system and then develop new works based on the generated suggestions. This growing trend of AI-assisted artistic production poses challenges in determining authorship. Consequently, the proliferation of ``semi-AI'' generated productions is increasing. In the music industry, this phenomenon may not be problematic, provided that AI is used judiciously \cite{louie2020novice,newman2023human}.

Therefore, we have focused our research on music melodies and lyrics that are fully generated by AI. While we acknowledge that partially AI-generated music, where a considerable portion of the melody is created by AI, can be detrimental to the music community by saturating the market, our research is advocating from a early research perspective to attract more research on this field. Determining the permissible degree of AI involvement in music production presents an intriguing area for future research. Additionally, our emphasis on the composition phase stems from our belief that it is the most critical stage in demonstrating creativity and ``musicality." This topic has been philosophically debated in the literature due to the inherently subjective nature of music. From an abstract perspective, music serves as a bridge for expressing an author's emotions and narratives through a combination of musical notes and lyrics \cite{supivcic1971expression}. Currently, AI struggles to generate authentic emotional expression \cite{li2025artificialemotionsurveytheories}. The remaining components, such as music mixing, primarily utilise AI in assembly and playback functions. Moreover, evaluating creativity and ``musicality" is challenging \cite{zhang2025aesthetics}. For instance, music talent shows often rely on audience votes rather than assessments of ``musicality'' to determine winners. Nonetheless, some theoretical frameworks exist for evaluating ``musicality," such as harmony consistency, albeit at a basic level \cite{elliott1987assessing}. We can also examine simpler features like melodic range and rhythmic entropy in our research. In response to these challenges, our goal is to create a comprehensive dataset encompassing a variety of music types and scenarios to further advance the field.

Similar to the well-established field of machine-generated text detection, which has benefited from widely used datasets like M4 \cite{wang2024m4}, we draw inspiration from these efforts and extend their ideas to the knowledge of music. This has led to the development of M6: a comprehensive database for Multi-Generator, Multi-Domain, Multi-Lingual and Cultural, Multi-Genre, and Multi-Instrument Machine-Generated Music Detection. We acknowledge that different generators produce distinct output distributions, resulting in various ``styles" of music, supporting by research indicating that watermarking techniques can effectively identify the outputs of different generators \cite{kirchenbauer2023watermark}. Consequently, we incorporate multiple generators into our dataset to enhance its diversity. Li et al.\ \cite{li2024audio} emphasise the two key stages of music production: the composition stage, where melody and harmony express the musical narrative, and the decoration stage, where sound design and mixing enhance the music's aesthetic quality. To reflect the diversity of composition, we have curated a multi-domain music collection suitable for various scenarios, such as background music or songs with lyrics, as well as multi-lingual and culturally diverse pieces tailored to different audiences. Additionally, for the decoration stage, selecting appropriate genres to shape the sound and choosing instruments to complement melodies are crucial steps in the process \cite{ma2024foundation}. These principles have guided the organisation of our dataset.

Our contributions are as follows:
\begin{itemize}
    \item We construct, to the best of our knowledge, the first comprehensive music database specifically designed for MGMD. We also provide detailed data analysis on the database, showing important intrinsic features. 
    \item We evaluate the performance of foundational detectors from various perspectives and we derive several key observations that can guide future research directions. Especially, Transformer-based models behaves slightly worse than CNN based feature extraction models.
    \item We will release our data and code publicly, and we plan to continuously update our repository by adding new generators, domains, and languages.
\end{itemize}

Therefore, the paper is organised as follows. In section \ref{sec:related}, we first review related work on MGMD databases. In section \ref{sec:process}, we then describe our database collection process and the design methodology behind it. section \ref{sec:analysis} presents basic data analysis, followed by an evaluation of the performance of foundational models in classification tasks in section \ref{sec:baseline}. Finally, in section \ref{sec:conclusion}, we summarise our findings, highlighting that the field is still evolving and calling for further research efforts.
\section{Music dataset}
\label{sec:related}
For existing datasets, we reference \textbf{FakeMusicCaps} \cite{comanducci2024fakemusiccaps} and \textbf{SONICS} \cite{rahman2024sonics} with a more detailed comparison of dataset provided in section \ref{sec:analysis}.

\subsection{AI-Generated Dataset}
\subsubsection{FakeMusicCaps} This dataset was designed to detect the authorship of the music. It is based on the MusicCaps dataset\footnote{https://paperswithcode.com/dataset/musiccaps}, from which human-annotated prompts were extracted and used to feed Text-to-Music models for generating MGM. The dataset includes five distinct music generators, including the autoregressive MusicGen \cite{copet2024simple} and four latent diffusion models. The performance of the dataset was evaluated using convolutional neural networks (CNNs) \cite{lecun2015deep}, though this approach is considered insufficient. 
\subsubsection{SONICS} This dataset was specifically designed for MGMD. It categorises music into three classes: fully fake, partially fake (distinguished by real music styles with fully fake), and real types of music songs. Lyrics are generated using LLMs, while melodies are produced by commercial tools such as Suno AI\footnote{https://suno.com/about}, which may present potential redistribution license issues. The dataset is evaluated using their proposed Spectro-Temporal Tokens Transformer (SpecTTTra) model.

\subsubsection{Importance of M6}
While existing datasets such as SONiCS and FakeMusicCaps have pushed the field of music authenticity detection forward, our proposed M6 dataset addresses a distinct and critical gap in the study of MGMD.

SONiCS is a large dataset focused on Synthetic Song Detection (SSD) with a goal of end-to-end AI-generated songs where vocals, music, lyrics, and style can all be generated. SONiCS is centered around long-range temporal modeling and introduces efficient models such as SpecTTTra. SONiCS is predominantly applied in the use of song-level authenticity detection, generally from consumer-facing websites like Suno and Udio, with high performance benchmarking but low regard for controlled diversity or analysis.

FakeMusicCaps, on the other hand, focuses on music-text alignment for tasks like fake caption detection or multimodal grounding, rather than standalone audio authenticity. It is thus not optimized for MGMD at the audio level.

In contrast, M6 is designed as a general-purpose, fine-grained benchmark for MGMD. It features diverse generation methods, including symbolic-to-audio and audio-to-audio models, not limited to end-to-end song synthesis platforms. Additionally, rich coverage across genres, cultures, instruments, and languages, making it more suitable for studying generalisation and cross-domain robustness. Moreover, comprehensive metadata and baseline results from multiple standard binary classifiers, supporting reproducibility and comparative research. All elements considered, M6 provides a foundation for rigorous, extensible research in MGMD essential for it, with distinctive contribution other than FakeMusicCaps and SONICS.

\subsection{Human-Made Dataset}
On the other hand, human-made music is essential in our databases, as it serves as positive samples. The vast availability of human-made music has grown significantly alongside the development of the music community. Several previous efforts have aggregated such music collections, including datasets like DALI \cite{meseguer2019dali}, MAESTRO \cite{hawthorne2018enabling}, and MusicNet \cite{thickstun2017learning}. Additionally, websites such as Spotify\footnote{https://open.spotify.com/}, NetEase Cloud\footnote{https://hr.163.com/product.html/music}, and Jamendo\footnote{https://www.jamendo.com}, provide a wealth of music, although their licensing requirements must be carefully considered. Based on our selection criteria, we have chosen the \textbf{GTZAN Dataset} \cite{GTZAN}, the \textbf{Free Music Archive (FMA)} \cite{defferrard2016fma}, \textbf{COSIAN} \cite{yamamoto2022analysis}, and the \textbf{Musical Instruments Sound Dataset (MISD)} \footnote{https://www.kaggle.com/datasets/soumendraprasad/musical-instruments-sound-dataset}, all of which are of high quality for inclusion in our research. The comparison of the dataset will be discussed in \ref{sec:analysis}.

\subsubsection{GTZAN}  
The GTZAN dataset is specifically organised by genre, featuring ten distinct categories.
It consists of 1,000 tracks, each with a duration of 30 seconds, sampled at 22.05 kHz. Originally designed for music genre recognition tasks, it remains a widely used benchmark
The dataset provides audio waveforms and spectrogram representations.

\subsubsection{FMA}  
The Free Music Archive (FMA) dataset encompasses a vast collection of freely accessible music tracks across various genres and styles. It is structured into different subsets for FMA-Small, Medium and Large based on the number of tracks and audio quality. Each track is encoded in MP3 format at 44.1 kHz and annotated with metadata, including genre labels, artist names, and album information. 

\subsubsection{COSIAN}  
The COSIAN dataset focuses on Japanese songs, specifically curated for music singing skill analysis. It includes annotations related to vocal quality, pitch accuracy, and expressive singing styles. The dataset is constructed using recordings from various sources, including platforms such as Spotify, and provides both raw audio and aligned lyrics. 
\subsubsection{MISD}  
The Musical Instrument Sound Dataset (MISD) includes high-quality recordings of four commonly used musical instruments: piano, guitar, violin, and flute. The dataset captures a wide range of playing techniques, dynamics, and articulations.

All these datasets are available under free licenses, making them valuable resources for research in music analysis and generation, especially for applications in MGMD detection.

\section{Database collection process}
\label{sec:process}

\begin{figure*}[htbp]
\centerline{\includegraphics[width=0.66\linewidth]{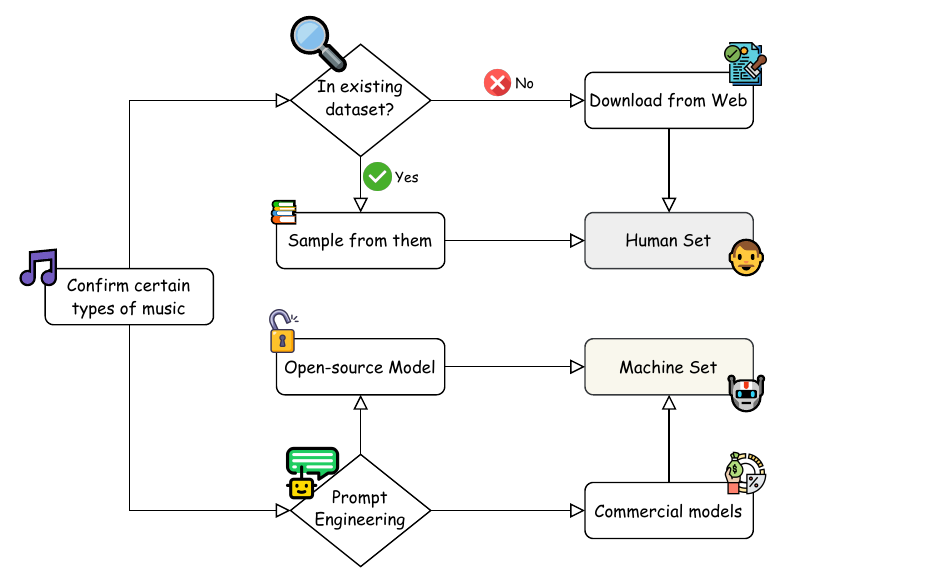}}
\caption{The process pipeline for collecting our database begins with determining the types of music to include. We aim to collect six key categories: music based on different instruments, different languages and cultures, with or without lyrics, various genres, differing music lengths, and general musics without any content restriction. Initially, we assess existing research datasets to see if they meet our requirements. If they do not, we obtain the necessary music from websites with appropriate licensing. Next, we generate our own set of MGM by utilising either research-based models or commercial models, conditioned on prompts generated by LLMs such as GPT-3.5.}
\label{fig:collection process pipeline}
\end{figure*}

Our data collection process is divided into two stages: reassemble human-made music and generating our own MGMs, as outlined in Figure \ref{fig:collection process pipeline}.

\subsection{M6 statistics overview}
In the creation of the M6 dataset, we focus on six distinct categories of music. The selected categories include: (a) different musical instruments, (b) diverse languages and cultural influences, (c) the presence or absence of lyrics (encompassing both vocal songs and instrumental background music across various domains), and (d) multiple musical genres. These classifications align with conventional categorisations outlined in prior surveys \cite{fu2010survey}.

Furthermore, we address a key limitation of existing audio processing models—their difficulty in handling long audio sequences. Prior research has shown that models such as Transformers struggle with capturing long-range dependencies in audio \cite{dong2020end}. To address this issue, we standardise the average duration of most music samples to 60 seconds by controlling the maximum length hyperparameters, thereby ensuring compatibility with contemporary models. Samples shorter than 60 seconds are retained in their original form, and additional instances of such samples are included to maintain a balanced overall duration distribution. Additionally, we introduce a dedicated subset (e) consisting of longer songs (2–3 minutes) to evaluate model robustness in processing extended sequences. For out-of-domain testing, we construct a random subset (f) comprising a diverse collection of general music. A summary of these resources, which are discussed in detail in the next two subsections, is provided in Table \ref{tab:datasets_resources}.
\begin{table}[htbp]
\caption{The resource of six types of music}
\begin{center}
\begin{tabular}{p{1cm}p{3.4cm}p{3.2cm}}
\hline
\textbf{Music} & \multicolumn{2}{p{6.6cm}}{\textbf{Resources}} \\
\cline{2-3}
\textbf{type} & \textbf{\textit{Human made}}& \textbf{\textit{MGM}} \\
\hline
(a)& MISD & AMG \& MG \& MGPT $^{\mathrm{a}}$\\
\hline
(b,c) $^{\mathrm{b}}$& NetEase Cloud for ZH; COSIAN (Spotify) for JP; Jamendo for EN$^{\mathrm{c}}$& Mureka  \\
\hline
(d)& GTZAN & AMG \& MG \& MGPT \\
\hline
(e)& Jamendo & Mureka \\
\hline
(f)& FMA & Soundraw \& AIVA \\
\hline

\multicolumn{3}{p{8cm}}{$^{\mathrm{a}}$ To differentiate between MGPT and MG, we use a smaller pre-trained MG as the base model for MGPT (\url{https://huggingface.co/facebook/musicgen-small}), while a larger pre-trained MG (\url{https://huggingface.co/facebook/musicgen-large}) serves as the MG model.}
\\
\multicolumn{3}{p{8cm}}{$^{\mathrm{b}}$ We merge these two types to focus on multilingual, lyric-based songs, ensuring that only one language appears in each song. Mixed-language international songs are excluded from our research. All other types, except for category (e), consist of background music.}
\\
\multicolumn{3}{p{8cm}}{$^{\mathrm{c}}$ The languages are ZH (Chinese), JP (Japanese), and EN (English). Except for this type, all others are based in English-speaking cultures.}
\vspace{-0.65cm}
\end{tabular}
\label{tab:datasets_resources}
\end{center}
\end{table}

\subsection{Human made music sampling}
Given the extensive availability of human-made music resources, we do not aim for exhaustive coverage. Instead, for song samples included in the datasets reviewed in section \ref{sec:related}, we randomly select a representative subset with a sufficient number of samples as part of M6 to reflect the distribution of features in human-composed music. To ensure compliance with copyright regulations, all selected datasets—most of which are reviewed in section \ref{sec:related}—are licensed for non-commercial use and redistribution. For datasets that do not fall under this category, we source samples from online platforms such as NetEase Cloud and Spotify, where usage is restricted to personal purposes; therefore, we provide only metadata in the supplementary materials. Notably, we manually review the selected songs to ensure that the recordings are not predominantly composed of interludes, thereby preserving the stylistic characteristics of the music.

\subsection{Machine generated music creation}
\subsubsection{Model selection}
While human-made music is widely available, MGM needs to be produced. Several types of music generation models exist, including Variational Autoencoder (VAE)-based Generators (e.\,g., MusicVAE \cite{roberts2018musicvae}), and Generative Adversarial Networks (GANs) for music (e.\,g., MuseGAN \cite{dong2018musegan}). Additionally, Transformer-based \cite{vaswani2017attention} Music Generators, such as MusicGen (MG) \cite{copet2024simple} and Awesome-Music-Generation (AMG) \cite{wei2024melody}, have demonstrated significant promise. There is also a growing body of work on conditional generation based on various musical features, such as style-based generation \cite{mao2018deepj}, lyrics-based generation \cite{lei2024songcreator}, and melody-based generation \cite{chen2020melody}. However, given the substantial power of LLMs, we have selected language model-based approaches for research and MGM generation, specifically AMG, MG, and MusicGPT (MGPT), an extended interface for MG.

AMG generates high-quality music by conditioning on melody inputs and utilising text prompts, effectively aligning melodies with text. However, it is limited to producing 10-second audio background music. MG instead utilises a language model conditioned on compressed discrete music representations and integrates prompt engineering for greater flexibility in music generation. The extended interface, \emph{MGPT}, enables the incorporation of additional music generators, expanding its functionality. However, both are limited to generating audio music with a maximum duration of 30 seconds.

We carefully crafted \textbf{prompts settings} for these language models. We utilise simple instruction-akin prompts for either research tools or commercial tools, while we utilise the lyrics generated by GPT-3.5 \cite{openai2023gpt35} with prompts \emph{Give me 100 30-second-{language} song lyrics.} for type (c), and \emph{Give 100 prompts to generate music.} for type (f) for Mureka. For example, the prompts for type(f) are more like \emph{Generate a 60-second song with a slow tempo, and subtle piano notes.} or \emph{Generate a 60-second song at a fast tempo, using guitar as the primary instrument.} The prompts are guided by principles of prompt engineering \cite{reynolds2021prompt}, which indicates that all prompts should be designed straightforward, concise, and easy to interpret, aiming to reduce the complexity of prompt understanding that could compromise the quality of MGM. Similarly, for AMG and MG, we use \emph{Please generate a song with {instrument / genres}}. Additionally, we introduce intentional randomness by varying prompts. For example, we use \emph{Give me {instrument based / genres} songs.} for MGT and \emph{One 30-second {language} song.} for Mureka (generating type (b, c)).

In addition to research-based models, commercial black-box tools also offer valuable capabilities for generating high-quality music, typically accompanied by purchasable licenses. Well-known commercial tools include Suno
(which explicitly states that the generated music cannot be used for AI research, so we omit this tool), AIVA\footnote{https://creators.aiva.ai/}, Soundraw\footnote{https://soundraw.io/}, Mureka\footnote{https://www.mureka.ai/}, and others. These tools are challenging to analyse comprehensively due to the proprietary nature of their technical architectures, which are often concealed for commercial reasons. While many similar tools exist, most are variants with minimal differences, primarily limited to user interfaces repackaging similar underlying models to optimise profitability. For this study, we selected the most relevant commercial tools based on usage rankings and Google search trends.

\subsubsection{Quality control}
Music generation, unlike other tasks, is a form of artistic production, which is inherently subjective and prone to bias when assessed by human annotators. To complement human judgment, we employed objective metrics to provide more consistent quality assessments for MGM and to ensure control over the generated data. Specifically, we selected Rhythmic Entropy \cite{dubnov2004rhythmic}, which measures the complexity of rhythmic patterns; Melodic Range \cite{huron2006sweet}, which captures the span between the lowest and highest pitches to reflect melodic consistency; and Harmonic-to-Noise Ratio (HNR) \cite{yegnanarayana1999harmonic}, which quantifies the proportion of harmonic components to noise, indicating the clarity and quality of harmony. Statistical analysis reveals that MGM and human-composed music differ significantly in rhythmic and melodic characteristics across most subsets, while HNR also shows significant differences in nearly all cases except for type f. Nevertheless, these numerical differences are relatively small, making them difficult to perceive in listening tests. This suggests that while future work could adapt such metrics as fine-grained objectives for further improving MGMD, the current MGM dataset already maintains high quality comparable to human-composed music. The detailed results are presented in Table \ref{tab:music_type_high_features}.

\begin{table}[t]
\renewcommand{\arraystretch}{1.2}
\caption{Analysis of M6 music quality control metrics across different subsets. 
Significance levels: * $p<0.05$, ** $p<0.01$, *** $p<0.001$, n.s. = not significant.}
\label{tab:music_type_high_features}
\centering
\begin{tabular}{@{}llccc@{}}
\hline
\textbf{Music Type} & \textbf{Source} & \textbf{Rhythmic Entropy} & \textbf{Melodic Range} & \textbf{HNR} \\ \hline
\multirow{2}{*}{a}  & AI    & 3.99 $\pm$ 0.741*** & 3295.11 $\pm$ 774.322*** & 7.91 $\pm$ 7.554*** \\
                    & human & 2.93 $\pm$ 1.549    & 3033.81 $\pm$ 1091.145   & 6.05  $\pm$ 14.377  \\ 
\multirow{2}{*}{bc} & AI    & 4.43 $\pm$ 0.458**  & 3742.16 $\pm$ 269.915*** & 10.54 $\pm$ 4.358*** \\
                    & human & 4.53 $\pm$ 0.382    & 3822.53 $\pm$ 101.302    & 7.89$\pm$2.717     \\ 
\multirow{2}{*}{d}  & AI    & 4.02 $\pm$ 0.719*** & 3450.75 $\pm$ 708.861*** & 5.86 $\pm$ 5.855* \\
                    & human & 4.70 $\pm$ 0.361    & 3790.52 $\pm$ 211.547    & 5.34 $\pm$ 5.005   \\ 
\multirow{2}{*}{e}  & AI    & 5.90 $\pm$ 0.448*** & 3783.75 $\pm$ 152.394*** & 11.84 $\pm$ 4.473*** \\
                    & human & 6.43 $\pm$ 0.337    & 3852.34 $\pm$ 5.073      & 6.79 $\pm$ 2.048   \\ 
\multirow{2}{*}{f}  & AI    & 5.05 $\pm$ 0.728*** & 3763.53 $\pm$ 247.817*** & 8.33 $\pm$ 4.297 (n.s.) \\
                    & human & 4.50 $\pm$ 0.921    & 3690.92 $\pm$ 442.422    & 8.52 $\pm$ 4.823   \\ \hline
\end{tabular}
\end{table}

\section{Database analysis}
\label{sec:analysis}

\subsection{Statistical analysis}
The novel dataset we developed demonstrates superior characteristics compared to existing datasets. Specifically, we evaluated our dataset against FakeMusicCaps and SONICS in terms of dataset size and scope, together with the usage of all existing human made music datasets, as summarised in Tables \ref{tab:dataset size} and \ref{tab:human data}. 

\begin{table}[t!]
\caption{A comparison of the M6 dataset size and scope with existing datasets.}
\label{tab:dataset size}
\centering
\scriptsize 
\begin{tabular}{@{}lllllll@{}}
\hline
\multicolumn{2}{c}{\textbf{Datasets}}       & \textbf{Pos. Count}$^{\mathrm{a}}$ & \textbf{Neg. Count}$^{\mathrm{a}}$ & \textbf{Total Count} & \textbf{Per Sec}$^{\mathrm{b}}$ & \textbf{Size} \\ \hline
\multirow{6}{*}{\textbf{M6}} 
& a                      & 1928                       & 6664                       & 8592                    & 19.31/18.1               & \multirow{5}{*}{N/A}             \\
& bc                     & 300                     & 300                     & 600                  & 28.83/30.19        &              \\
& d                      & 1000                    & 1650                    & 2650                 & 30.02/18.13        &            \\
& e                      & 71                      & 80                      & 151                  & 203.85/154.14      &             \\
& f                      & 1000                    & 500                     & 1500                 & 29.99/58.06        &            \\
& \textbf{Total}         & 4299                       & 9194                      & 13493                   & 21.6                  & 28.3GB \\ \hline
\multicolumn{2}{l}{\textbf{FakeMusicCaps}}  & 5521                    & 27605                  & 33126                & 10                & 16.3GB        \\ \hline
\multicolumn{2}{l}{\textbf{SONICS}}         & 48,090                  & 49,074                 & 97,164               & 176.03            & 32.8GB          \\ \hline
\multicolumn{7}{p{8cm}}{$^{\mathrm{a}}$ `Pos./Neg. Count' refers to the number of positive and negative cases, categorised as human-generated or AI-generated instances.}
\\
\multicolumn{7}{p{8cm}}{$^{\mathrm{b}}$ `Per Sec' indicates the number of cases per second averagely, with the left value representing positive cases and the right value representing negative cases. }

\vspace{-0.3cm}

\end{tabular}
\end{table}

\begin{table}[t!]
\caption{Human made dataset comparison}
\label{tab:human data}
\centering
\begin{tabular}{@{}p{1cm}p{0.6cm}p{0.8cm}p{1cm}p{4cm}@{}}
\hline
\textbf{Datasets} & \textbf{Count} & \textbf{Per Sec} & \textbf{Selected Count} & \textbf{Comments} \\ \hline
MISD & 2628 & 19.3 & 1928 & Only for Instruments Recognition \\ \hline
FMA                              & 8000 & 30   & 1000 & GTZAN alike                \\\hline
COSIAN                           & 168  & N/A$^{\mathrm{a}}$  & 100  & Japanese original songs    \\\hline
GTZAN                            & 1000 & 30   & 1000 & Only for Genres Recognition \\ \hline
\multicolumn{5}{p{8cm}}{$^{\mathrm{a}}$ They are all original Japanese songs.}
\vspace{-0.65cm}

\end{tabular}
\end{table}

Our dataset encompasses a broader scope and supports more diverse applications compared to existing datasets shown above. While the number of cases for MGM exceeds those for human-generated instances, deliberate biases were introduced in type (a) to facilitate robustness testing. For all other types, efforts were made to ensure temporal balance in total. Consequently, the M6 dataset is both comprehensive and well-balanced, addressing a wider range of research scenarios.

We also analyse some physical features of our collected data to represent the quality of our collected data and provide simple feature overviews in Table \ref{tab:music_type_features}.

These four features are derived from the time-frequency representation of audio signals and reflect their underlying physical characteristics. Spectral Centroid (SC) \cite{peeters2004large} and Spectral Bandwidth (SB) \cite{tzanetakis2002musical} show significant differences between MGM and human-composed music across several subsets, with MGM generally exhibiting lower centroid values and narrower bandwidths. This indicates reduced spectral variation and lower perceived brightness in MGM compared to human compositions. Zero Crossing Rate (ZCR) \cite{kumar2011zero}, which measures the frequency of sign changes in the signal, displays mixed results: significant differences appear in subsets a, e, and f, but not in bc and d. Root Mean Square Energy (RMSE) \cite{allen1994rms}, which captures the signal’s magnitude, shows consistently significant differences, with MGM tending to have lower energy levels than human-composed music. Collectively, these findings suggest that MGM differs systematically from human music in its spectral properties, while temporal features such as ZCR exhibit variability across subsets. These insights highlight potential directions for refining feature design to improve MGMD. 

\begin{table}[h!]
\renewcommand{\arraystretch}{1.2}
\caption{Comparison of M6 music physical features for different subsets, including spectral centroid (SC), spectral bandwidth (SB), zero-crossing rate (ZCR), root mean square energy (RMSE). 
Significance levels: * $p<0.05$, ** $p<0.01$, *** $p<0.001$, n.s. = not significant.}
\label{tab:music_type_features}
\centering
\begin{tabular}{@{}llcccc@{}}
\hline
\textbf{Music Type} & \textbf{Source} & \textbf{SC} & \textbf{SB} & \textbf{ZCR} & \textbf{RMSE} \\ \hline
\multirow{2}{*}{a}  & AI    & 1917.56 $\pm$ 1145.688*** & 1994.77 $\pm$ 833.788*** & 0.08 $\pm$ 0.057*** & 0.13  $\pm$ 0.057*** \\
                    & human & 2592.89 $\pm$ 2643.877    & 2726.77 $\pm$ 1627.095   & 0.07 $\pm$ 0.108    & 0.05  $\pm$ 0.051    \\ 
\multirow{2}{*}{bc} & AI    & 2803.62 $\pm$ 1204.083**  & 3290.99 $\pm$ 785.322*** & 0.06 $\pm$0.032 (n.s.) & 0.12 $\pm$0.022 (n.s.) \\
                    & human & 2603.67 $\pm$ 582.072     & 2772.31 $\pm$ 529.082    & 0.06 $\pm$0.021     & 0.12 $\pm$0.090    \\ 
\multirow{2}{*}{d}  & AI    & 2147.78 $\pm$ 976.749 (n.s.) & 2246.54 $\pm$768.428 (n.s.) & 0.10 $\pm$0.055 (n.s.) & 0.15 $\pm$0.068*** \\
                    & human & 2202.42 $\pm$ 715.718     & 2242.76 $\pm$526.253     & 0.10 $\pm$0.042     & 0.13 $\pm$0.066    \\ 
\multirow{2}{*}{e}  & AI    & 2172.73 $\pm$ 914.650***  & 2877.98 $\pm$677.504*** & 0.04 $\pm$0.022**   & 0.11 $\pm$0.017*** \\
                    & human & 2793.74 $\pm$ 565.553     & 3302.83 $\pm$406.052     & 0.05 $\pm$0.016     & 0.22 $\pm$0.051    \\ 
\multirow{2}{*}{f}  & AI    & 2427.49 $\pm$ 1143.858 (n.s.) & 2818.93 $\pm$941.396 (n.s.) & 0.06 $\pm$0.035*** & 0.08 $\pm$0.029*** \\
                    & human & 2354.05 $\pm$ 1104.776    & 2860.12 $\pm$973.042     & 0.05 $\pm$0.032     & 0.18 $\pm$0.097    \\ \hline
\end{tabular}
\end{table}

\begin{figure*}[t]
\centering
\includegraphics[width=\textwidth]{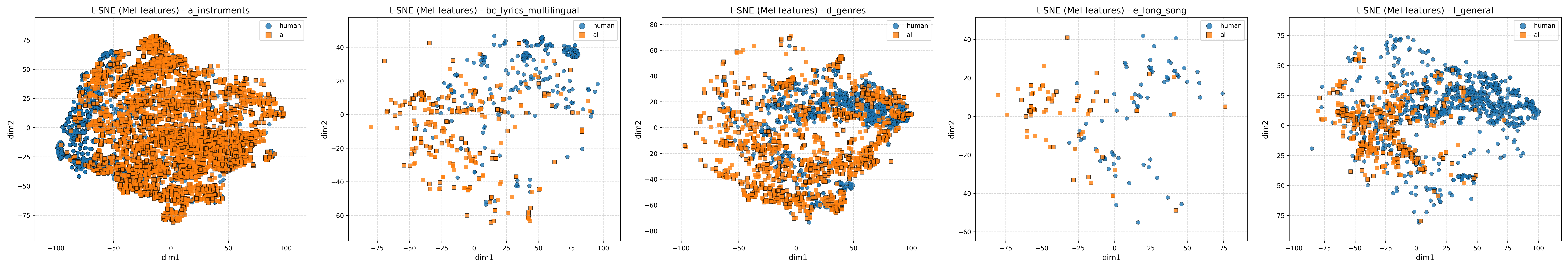}
\caption{T-SNE visualisation for Mel-spectrogram features. The curves for three models are shown from left to right: (a), (b, c), (d), (e), and (f).}
\label{fig:macro-af}
\end{figure*}

\begin{figure}[t]
\centering
\includegraphics[width=0.5\textwidth]{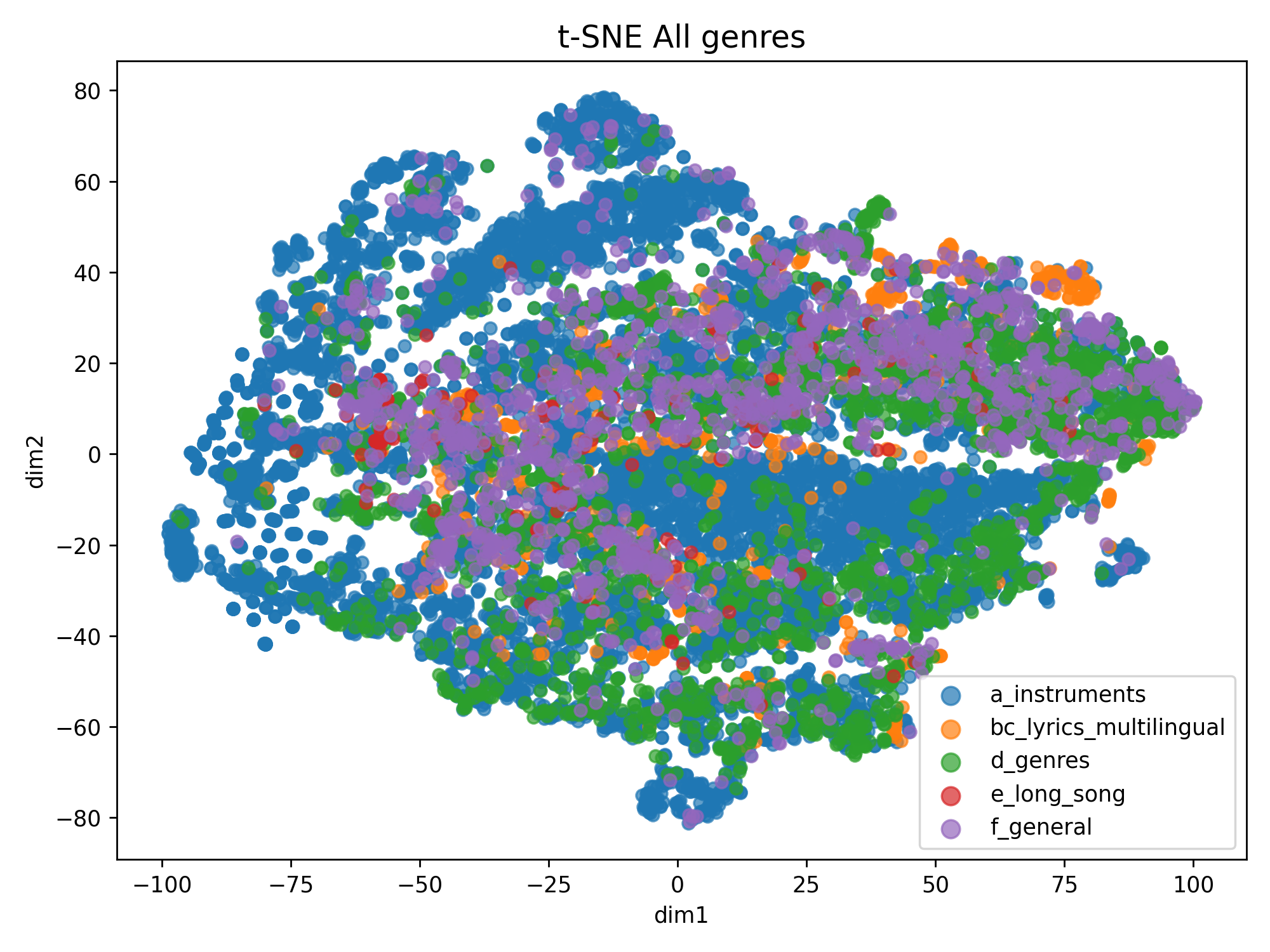}
\caption{T-SNE visualisation for Mel-spectrogram features of the whole M6 dataset.}
\label{fig:macro-all}
\end{figure}

In addition to the statistical analyses conducted to examine differences between Machine-generated and human-composed samples, we further investigate these distinctions from a machine learning perspective by analysing the distribution of audio representations within the feature space. Macro-level visualisations of AI and human samples for each music subtype in M6, as well as aggregated samples from the entire dataset, are presented in Figures \ref{fig:macro-af} and \ref{fig:macro-all}. These visualisations were generated using t-SNE \cite{maaten2008visualizing} applied to Mel-spectrogram features. The results suggest that there is no clear decision boundary between subtypes, and substantial overlap exists among data points, highlighting the inherent difficulty of this classification task. Moreover, the overall distribution indicates that overlaps within subtypes are expected, as musical categories are not strictly disjoint. For example, a guitar music sample may also be categorised as pop music. Nonetheless, each subtype exhibits distinctive clusters. For instance, a visible grouping of blue points in the upper-left region of the figure, demonstrating both the unique distributional characteristics of each subtype and the necessity of accounting for such diversity in the analysis.
For a more quantitative analysis beyond simple visualization, we computed the intra-feature L2 distances of each subtype in the embedding space for AI samples, human samples, and AI–human comparisons, as presented in Table~\ref{tab:distant}. The distances were calculated as the average pairwise values between points in the embedding space. On average, the difference between AI and human samples is noticeable, suggesting the potential for distinguishing between them. However, the classification of individual cases remains a challenge for the task for further discussion.
\begin{table}[h]
\centering
\begin{tabular}{lccc}
\hline
Type & AI Sample & Human Sample & AI vs Human Samples\\
\hline
a & 144.897 & 184.325 & 234.133 \\
bc & 126.484 & 111.540 & 153.759 \\
d & 150.877 & 117.471 & 139.488 \\
e & 117.837 & 75.673 & 150.846 \\
f & 110.569 & 138.210 & 164.795 \\
\hline
\end{tabular}
\caption{Intra distance comparison between AI, Human, and AI vs Human samples.}
\label{tab:distant}
\end{table}

In addition to the intra-distance analysis, we also measured the distributional differences of each subtype using L2-distance on Mel-spectrogram embeddings. For each subtype, we computed the centroid by averaging all AI or human samples, which served as the representative point. And we calculate distance based on the representative points. The resulting distances are reported in Table~\ref{tab:inter_comparison}. The results indicate that human samples, particularly in subtypes (a), (bc), and (d), exhibit substantial variation, whereas subtype (f) appears to cover many samples as this out-of-domain subtype is not specially curated to show certain music type. Moreover, AI samples are generally more similar to each other than human samples. We attribute this to the fact that machine-generated data are explicitly trained to mimic human samples, which may lead to the emergence of universal patterns across generators. In contrast, human data are inherently more diverse, with a larger number of individual creators contributing to variability. These observations suggest future research directions in identifying and characterising the shared patterns learned by machine generators.

\begin{table}[ht]
\centering
\begin{tabular}{lccccc}
\hline
Type & a & bc & d & e & f \\
\hline
a        & - & 27.000 & 39.920 & 92.013 & 62.232 \\
bc & 256.487 & - &  52.726 & 75.302 & 48.184  \\
d             & 244.232 & 33.954  & - & 125.455 & 96.891   \\
e         & 206.966 & 57.517  & 38.224  & - & 36.342  \\
f            & 228.497 & 42.006  & 23.500  & 27.274  & - \\
\hline
\end{tabular}
\caption{Intra distance comparison between AI, Human.  
\textbf{Upper triangle: AI}, \textbf{Lower triangle: Human}.}
\label{tab:inter_comparison}
\end{table}

\subsection{Human assessment}

\begin{figure*}[h]
\centering
\includegraphics[width=0.95\textwidth]{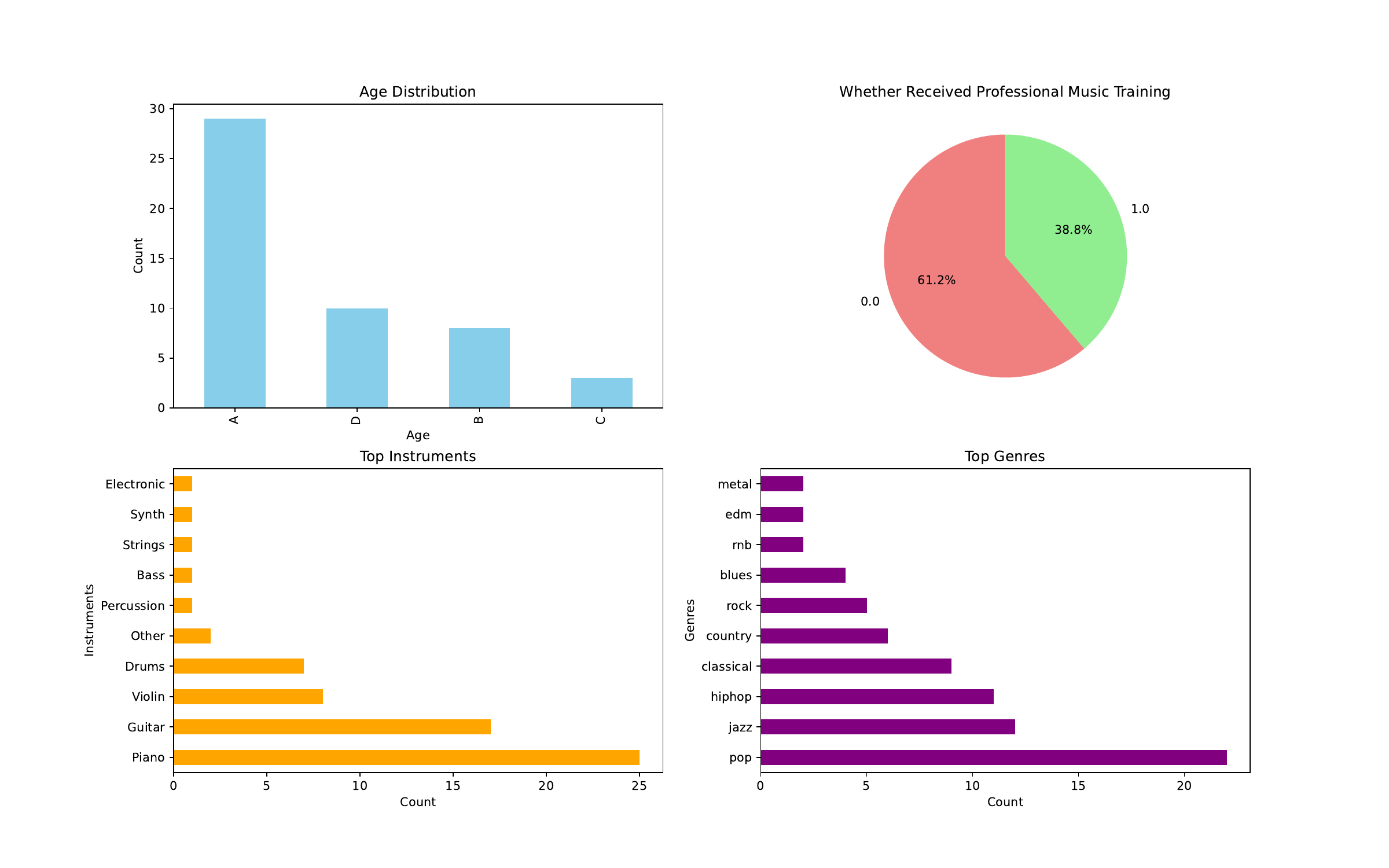}
\caption{Demographic distribution of the 50 participants, including age groups (A: 18–25, B: 26–35, C: 36–45, D: 46+), history of professional music training, frequency of exposure to musical instruments, and preferred music genres.}
\label{fig:demographics}
\end{figure*}

As music is a form of art and inherently subjective, in addition to statistical analysis, we invited 50 participants to evaluate this dataset. All invitations adhered to ethical regulations, and all participants provided informed consent to share their perspectives for research purposes. The demographics of the participants are presented in Figure \ref{fig:demographics}. Efforts were made to ensure that the invited participants were as demographically balanced as possible.

\begin{figure*}[h]
\centering
\includegraphics[width=0.95\textwidth]{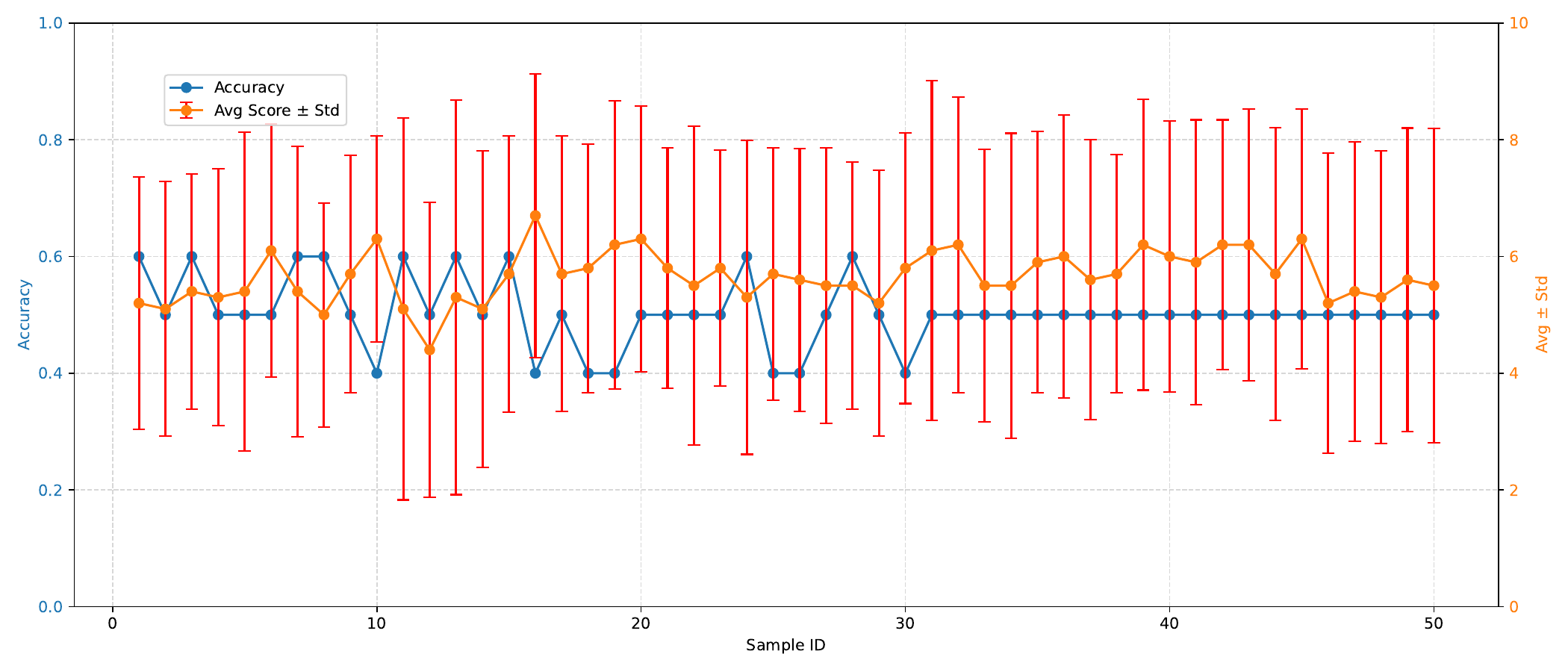}
\caption{Human annotation of 50 samples. We report the accuracy of human judgments and the average melodiousness score (1–10) for each sample, along with the standard deviation.}
\label{fig:human_assess}
\end{figure*}

We asked participants to judge whether the presented samples were MG or human-made, and to rate the degree of melodiousness of each music clip. We randomly sampled 50 music pieces from M6, ensuring balanced representation across subtypes (a) to (f), with 25 pieces from MD and 25 from human-made sources. Each participant was assigned 10 music pieces (5 machine generated cases and 5 human made cases) to annotate, resulting in 10 reviews per piece. However, they were not told about the number of human made cases they received to avoid inadvertent hint on their subjective assessment. Additionally, they were introduced about MGM and given samples from Suno \footnote{\url{https://suno.com/}}. However, they were explicitly informed that MGMs differ from such systems and were advised not to base their expectations solely on those examples. The annotation results are presented in Figure \ref{fig:human_assess}. The overall average accuracy across all samples was 0.504, indicating that our dataset is highly challenging for humans to distinguish.
\section{Baseline model result and discussion}
\label{sec:baseline}
We implement models on the M6 dataset to establish baseline performance scores. This allows for a systematic evaluation of how well existing models handle the dataset’s diverse characteristics. These benchmark scores serve as a reference for future improvements and comparisons.

\subsection{Music detectors selection}
\subsubsection{Experiment setup}
The MGMD is a binary classification problem for classifying whether it is generated by machine or human. The existing models for this task are limited, with very few detectors currently available and under development. Furthermore, these models lack a sufficient baseline, and their approaches are highly fragmented \cite{li2024audio}. However, drawing inspiration from binary classification and deepfake audio detection tasks, we select five representative types of classification models for evaluation: Quadratic Support Vector Machine (Q-SVM) \cite{singh2021detection} as a traditional machine learning model, ResNet \cite{resnet2015} as a deep neural network model, ViT \cite{dosovitskiy2020image} as a Transformer-based model, and Wav2Vec 2.0 \cite{baevski2020wav2vec20frameworkselfsupervised} and HuBERT \cite{hsu2021hubertselfsupervisedspeechrepresentation} as audio processing models. These models are selected based on their frequent adoption in deepfake audio detection tasks.

\emph{Q-SVM} is a variant of SVM that utilises a quadratic kernel function to map input data into a higher-dimensional space, which is a simple, yet effective classifier, while \emph{ResNet} is a CNN that incorporates residual connections to prevent information loss and facilitate the training of deeper networks. This model excels at extracting latent image features. \emph{ViT} is a transformer-based model that treats images as sequences of patches, leveraging the self-attention mechanism to capture intricate patterns and dependencies. \emph{Wav2Vec 2.0} and \emph{HuBERT} could accept raw music as input to analyse features with pre-trained processor.

We split the dataset into training, validation, and test sets for each type, with proportions of 0.8×0.8, 0.8×0.2, and 0.2, respectively. Notably, we keep the same data split set for fair experiments comparison. To facilitate subsequent analysis, all audio files in WAV format were processed using the \emph{librosa} library \cite{mcfee2015librosa} to extract Mel-spectrogram representations, with the exception of Wav2Vec 2.0 and HuBERT, which operate directly on raw audio, and Q-SVM, which relies on spectral statistics as input. The hyperparameters are shown in Table \ref{tab:hyperparameters}. \begin{table}[ht]
\centering
\caption{Hyperparameter Settings}
\label{tab:hyperparameters}
\begin{tabular}{l|c}
\hline
\textbf{Hyperparameter} & \textbf{Value} \\ \hline
Batch Size             & 64              \\
Epochs       & 10             \\
GPU & Tesla V100 \\
input size & 224 $\times$ 224 \\
Learning Rate          & 0.001          \\
optimiser & Adam \\\hline
\end{tabular}
\end{table}
For the ResNet18 architectures, we adopt the default values specified in their original papers. For the ViT model, we use the pre-trained \emph{google/vit-base-patch16-224} and adapt the input dimension from 3-channel images to a single channel to accomodate our input. For Q-SVM, we set the kernel to `poly,' the degree to 2, and the coefficient $c_0$ to 1.

\subsubsection{Results}
We evaluate the aforementioned models through accuracy, and F1 score shown in Table \ref{tab:model_comparison}, \ref{tab:model_comparison_outofdomain}.

\begin{table*}[h]
\centering
\caption{Performance Comparison of Accuracy and F1 score of Models Across Different Music Types}
\label{tab:model_comparison}
\begin{tabular}{ccccccccccc}
\hline
\multirow{2}{*}{\textbf{Music Type}} & \multicolumn{2}{c}{\textbf{Q-SVM}} & \multicolumn{2}{c}{\textbf{ResNet18}} & \multicolumn{2}{c}{\textbf{ViT}} & \multicolumn{2}{c}{\textbf{Wav2Vec 2.0}} & \multicolumn{2}{c}{\textbf{HubBERT}} \\ \cline{2-11} 
                                      & \textbf{Acc} & \textbf{F1}  & \textbf{Acc} & \textbf{F1}  & \textbf{Acc} & \textbf{F1}& \textbf{Acc} & \textbf{F1}& \textbf{Acc} & \textbf{F1}  \\ \hline
a                       & .982         & .957         & .974         & .974         & .985         & .985    &.787&.693&   .966&.966 \\ 
bc                      & .917         & .915         & .883         & .883         & .667         & .630 &.758&.757   &.858&.856     \\ 
d                       & .975         & .967         & .992         & .992         & .951         & .950     &.638&.497 &.796&.786   \\ 
e                       & .968         & .966         & .903         & .901         & .548         & .388    &.548&.388   &.645&.629  \\ 
abcde                     & .959            & .923           &  .973           & .973            & .949            & .949 &.736&.625   &.889&.885       \\ \hline

\end{tabular}
\end{table*}

\begin{table*}[h]
\centering
\caption{Out of Domain Performance Comparison evaluation on (f) of Models Across Different Music Types}
\label{tab:model_comparison_outofdomain}
\begin{tabular}{ccccccccccc}
\hline
\multirow{2}{*}{\textbf{Music Type}} & \multicolumn{2}{c}{\textbf{Q-SVM}} & \multicolumn{2}{c}{\textbf{ResNet18}} & \multicolumn{2}{c}{\textbf{ViT}} & \multicolumn{2}{c}{\textbf{Wav2Vec 2.0}} & \multicolumn{2}{c}{\textbf{HubBERT}} \\ \cline{2-11} 
                                      & \textbf{Acc} & \textbf{F1}  & \textbf{Acc} & \textbf{F1}  & \textbf{Acc} & \textbf{F1}& \textbf{Acc} & \textbf{F1}& \textbf{Acc} & \textbf{F1}  \\ \hline
a                                     & .102         & .061        & .495         & .484         & .162         & .111  & .333 & .167 & .261 & .149       \\ 
bc                                    & .766         & .828         & .679         & .621         & .665         & .610  &.651 & .660 & .506 &.498       \\ 
d                                     & .745        & .767        & .803         & .808         & .508         & .471   &.333&.167&.503&.491      \\ 
e                                     & .625         & .733         & .417         & 0.415         & .333         & .167    &.333&.167&.545&.556     \\ 
abcde                                   & .420           & .565            & .635            & .645           & .491           & .504 &.333&.167&.315&.288           \\ \hline

\end{tabular}
\end{table*}

It can be observed that ResNet-18 achieves the best performance in standard cases, demonstrating a relatively high level of accuracy and F1 score for the detection task. While Q-SVM and ViT perform marginally worse, their results remain competitive. Wav2Vec 2.0 performs worse, as it is not specifically pre-trained for this task. HuBERT also performs well but does not surpass ResNet-18. These transformer-based architectures indicate that, although they prioritise contextual learning, such an approach may not be necessary for this task, which likely requires stronger spatial feature extraction capabilities. Furthermore, the similarity between the accuracy and F1-score is supporting the claim that our dataset is relatively balanced. Moreover, it demonstrates that long-term music (e) presents the most significant challenge, advocating more models to deal with the dependency in the process. Overall, the models exhibit fundamentally strong performance in the in-domain evaluation.

\begin{figure*}[h]
\centering
\includegraphics[width=0.75\textwidth]{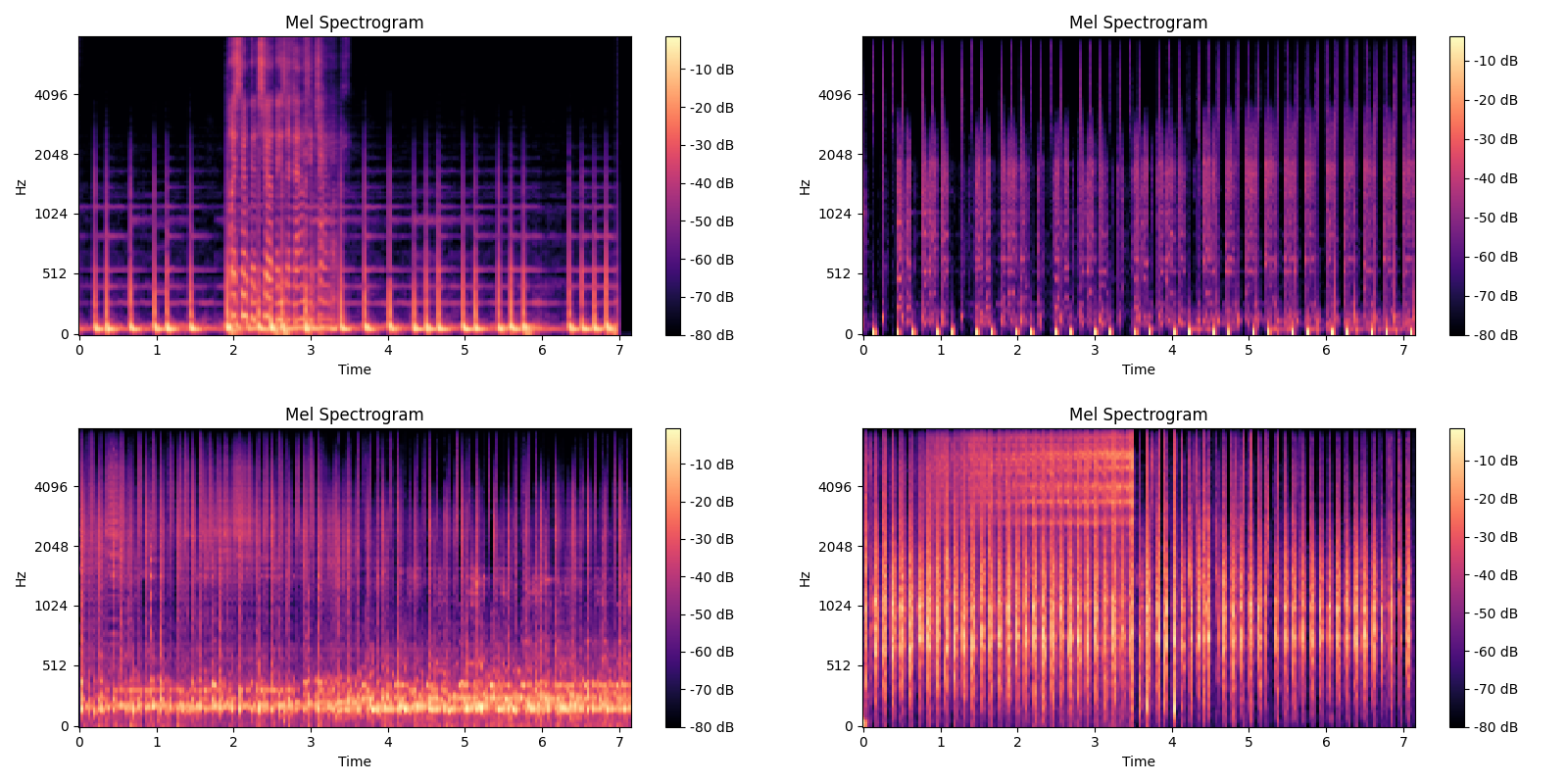}
\caption{Mel-spectrogram of misclassified samples in type (a) with ResNet18 (up) and ViT (down).}
\label{fig:case_compare_transvsres}
\end{figure*}

We analysed several misclassified cases in subtype (a) to compare the best-performing convolutional model, ResNet18, with a transformer-based model, for which we selected ViT as a representative. These were further contrasted with correctly classified samples, as shown in Figure~\ref{fig:case_compare_transvsres}. Our observations suggest that ResNet18 tends to misclassify samples with pronounced high-frequency components. We attribute this to the fact that ResNet treats the Mel-spectrogram primarily as an image input, where excessive high-frequency information may blur the relative positional relationships within the representation. In contrast, ViT is more likely to misclassify cases where the temporal sequence lacks prominent distinguishing features. This supports our earlier claim that transformer-based models focus more heavily on contextual modeling. Nevertheless, we emphasize that such interpretations should be substantiated with more advanced explainable AI methods, which we advocate as a promising direction for future research.

\paragraph{Comparison with SOTA audio deepfake detection and ensemble models}

To better evaluate the quality of our dataset and check whether music could directly use audio deepfake detection model. We select RawNet \cite{jung2019rawnetadvancedendtoenddeep}, Audio Anti-Spoofing using Integrated Spectro-Temporal Graph Attention Networks (AASIST) \cite{jung2021aasistaudioantispoofingusing}, and Audio Spectrogram Transformer (AST) \cite{gong2021astaudiospectrogramtransformer} as representatives. Also, we select XGBoost \cite{Chen_2016} as ensemble models for comparison as well. The results are shown in Table \ref{tab:model_comparison_audio} and \ref{tab:model_comparison_audio_outofdomain}. 

\begin{table*}[h]
\centering
\caption{Performance Comparison of Accuracy and F1 score of SOTA Models in Audio Deepfake Detection and Ensemble Models Across Different Music Types}
\label{tab:model_comparison_audio}
\begin{tabular}{ccccccccccc}
\hline
\multirow{2}{*}{\textbf{Music Type}} & \multicolumn{2}{c}{\textbf{RawNet}} & \multicolumn{2}{c}{\textbf{AASIST}} & \multicolumn{2}{c}{\textbf{AST}} & \multicolumn{2}{c}{\textbf{XGBoost}} \\ \cline{2-9} 
& \textbf{Acc} & \textbf{F1} & \textbf{Acc} & \textbf{F1} & \textbf{Acc} & \textbf{F1} & \textbf{Acc} & \textbf{F1} \\ \hline
a      & .839 & .852 & .978 & .978 & .787 & .693 & .986 & .986 \\ 
bc     & .850 & .850 & .733 & .713 & .483 & .315 & .842 & .839 \\ 
d      & .979 & .979 & .966 & .966 & .625 & .480 & .960 & .960 \\ 
e      & .645 & .574 & .710 & .671 & .452 & .281 & .936 & .935 \\ 
abcde  & .957 & .957 & .945 & .944 & .726 & .611 & .970 & .970 \\ \hline
\end{tabular}
\end{table*}

\begin{table*}[h]
\centering
\caption{Out-of-Domain Performance Comparison on (f) of SOTA Models in Audio Deepfake Detection and Ensemble Models Across Different Music Types}
\label{tab:model_comparison_audio_outofdomain}
\begin{tabular}{ccccccccccc}
\hline
\multirow{2}{*}{\textbf{Music Type}} & \multicolumn{2}{c}{\textbf{RawNet}} & \multicolumn{2}{c}{\textbf{AASIST}} & \multicolumn{2}{c}{\textbf{AST}} & \multicolumn{2}{c}{\textbf{XGBoost}} \\ \cline{2-9} 
& \textbf{Acc} & \textbf{F1} & \textbf{Acc} & \textbf{F1} & \textbf{Acc} & \textbf{F1} & \textbf{Acc} & \textbf{F1} \\ \hline
a      & .605 & .558 & .187 & .111 & .333 & .167 & .020 & .017 \\ 
bc     & .651 & .659 & .523 & .494 & .667 & .533 & .718 & .726 \\ 
d      & .758 & .763 & .761 & .766 & .333 & .167 & .813 & .819 \\ 
e      & .378 & .263 & .564 & .546 & .667 & .533 & .029 & .029 \\ 
abcde  & .410 & .427 & .445 & .454 & .333 & .167 & .303 & .317 \\ \hline
\end{tabular}
\end{table*}

It could be told that the three models designed for audio deepfake detection and ensemble models XGBoost, demonstrate overall stable performance on the music task. Among audio deepfake detection models, AASIST shows the best performance, achieving above 0.94 in both accuracy and F1-score across especially in subtype a and d, indicating strong classification consistency and robustness. RawNet follows closely, particularly achieving near-perfect recognition on categories a and d, suggesting that its convolutional acoustic feature extraction is still effective at capturing temporal-frequency patterns. In contrast, AST exhibits relatively lower performance, especially in the out-of-domain condition, indicating its feature modeling is sensitive to input distribution shifts. XGBoost, the ensemble model, maintains stable accuracy across multiple categories, demonstrating the competitiveness of traditional tree-based models when provided with structured acoustic features.

Compared with the baseline models (Q-SVM, ResNet18, ViT, Wav2Vec2.0, HubERT), these audio deepfake models do not show a considerable overall advantage. For instance, while AASIST and RawNet slightly outperform in-domain accuracy, their results are comparable to ResNet18; in the out-of-domain test, AASIST exhibits slightly better generalization than HubERT but remains at a similar performance level overall. The comparable results suggest that though music and audio share similar features, the architectural advantages of speech deepfake models do not directly transfer to music as there are no explicit performance gain over traditional models. Compared to speech, music signals have more complex pitch variations, polyphonic structures, and broader spectral dynamics, which reduce the stability of discriminative features learned by models optimized for speech. Therefore, music audio deepfake detection may require specifically redesigned acoustic front-ends and temporal modeling mechanisms to better capture the multi-layered structural characteristics of music and adapt to domain distribution differences.
\subsubsection{Out of domain test analysis}
When evaluating models trained on each type using (f) as an out-of-domain test set, performance across all models and types is 
considerably poor. These results highlight the substantial challenges posed by the dataset and underscore the intrinsic complexity of the task. Furthermore, the findings indicate that the models lack robustness and fail to adequately capture underlying musicological characteristics. In practical applications of MGMD, encountering novel, out-of-domain music is inevitable. This underscores the urgent need for more robust models capable of generalising across diverse musical contexts, alongside models with improved explainability to enhance their practical utility.

After observing the performance degradation on the out-of-domain test with subtype (f), we conducted an analysis to better understand the mechanisms and weaknesses underlying model generalization failures. We first designed an additional experiment in which ResNet and ViT were trained on subtype (a) and evaluated on subtypes (bc) and (d), which are considered out-of-domain. Furthermore, we carried out a more fine-grained analysis across different languages and genres to identify which types of music are more prone to misclassification under cross-domain conditions. The results are presented in Table~\ref{tab:accuracy_subgroup}.

\begin{table}[h]
\centering
\begin{tabular}{lllc}
\hline
Task & Model & Subgroup & Accuracy \\
\hline
 \multirow{6}{*}{bc} 
 & \multirow{3}{*}{ResNet} & Chinese  & 0.500 \\
 &    & English  & 0.485 \\
 &    & Japanese & 0.465 \\
 &   \multirow{3}{*}{ViT} & Chinese  & 0.530 \\
 &    & English  & 0.535 \\
 &    & Japanese & 0.445 \\
\hline
 \multirow{22}{*}{d} 
 & \multirow{11}{*}{ResNet} & blues     & 0.121 \\
 &    & classical & 0.155 \\
 &    & country   & 0.185 \\
 &    & disco     & 0.219 \\
 &    & hiphop    & 0.257 \\
 &    & jazz      & 0.224 \\
 &    & metal     & 0.140 \\
 &    & pop       & 0.377 \\
 &    & reggae    & 0.245 \\
 &    & rock      & 0.177 \\
\cline{2-4}
 &   \multirow{11}{*}{ViT} & blues     & 0.381 \\
 &    & classical & 0.377 \\
 &    & country   & 0.374 \\
 &   & disco     & 0.381 \\
 &    & hiphop    & 0.396 \\
 &    & jazz      & 0.383 \\
 &    & metal     & 0.355 \\
 &    & pop       & 0.385 \\
 &    & reggae    & 0.396 \\
 &    & rock      & 0.374 \\
\hline
\end{tabular}
\caption{Out of domain test accuracy of ResNet and ViT across subgroups for types (bc, d).}
\label{tab:accuracy_subgroup}
\end{table}
In this out-of-domain evaluation setting, we observe substantial variation in model performance across linguistic and stylistic subgroups. Within the multilingual lyrics domain, Japanese samples yield the lowest detection accuracy, whereas Chinese and English samples perform comparatively better. This suggests that linguistic factors such as phonetic structure and rhythmic patterns introduce domain-specific variability, as Japanese is primarily an agglutinative language characterised by distinct rhythmic properties. Furthermore, the relative scarcity of Japanese music data compared to Chinese and English may also contribute to the observed performance gap.

\begin{figure*}[h]
\centering
\includegraphics[width=0.75\textwidth]{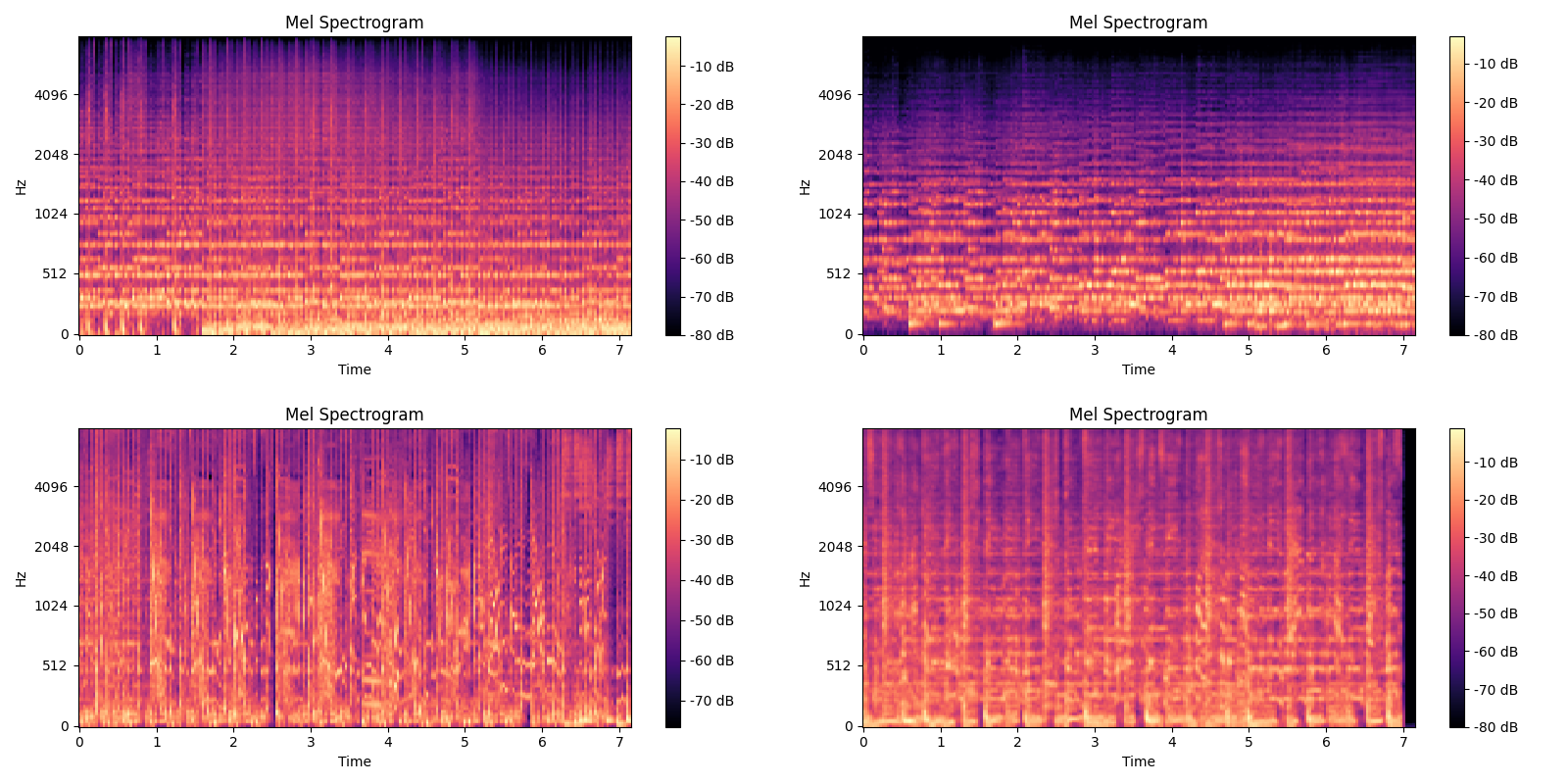}
\caption{Mel-spectrogram sample visualisation of out of domain test of music genres. Left-up(classic correct), Left-down(pop correct), Right-up(classic wrong), Right-down(pop wrong)}
\label{fig:case_ood}
\end{figure*}

For music style subgroups, ResNet achieves relatively low accuracy on classical, metal, and rock, while genres such as pop, hip-hop, and reggae are detected with higher accuracy. This suggests that musical styles with complex structures and diverse timbres are more difficult for the model to generalize, whereas styles with stronger rhythmic regularities or more repetitive acoustic patterns transfer more effectively. To better interpret these errors, we present misclassified samples from ResNet in Figure \ref{fig:case_ood}. The results indicate that the model tends to misclassify cases with broadly similar acoustic tendencies, particularly when the music lacks distinctive genre-specific features. For example, classical music often contains frequent pauses and chordal textures, leading to intermittent patterns in the mel-spectrogram (as shown in the upper-right example), while pop music typically exhibits more regular rhythmic patterns (as in the lower-right example). ResNet fails to capture these genre-specific cues, which contributes to misclassification. These observations, however, warrant further validation through more advanced explainable AI techniques.

Comparing model architectures, the transformer-based ViT consistently achieves higher detection accuracies than ResNet across both linguistic and genre subgroups. This highlights the advantage of contextual modeling in transformers, which appears to provide greater robustness against distributional shifts.

Additionally, we conducted a systematic analysis of the f-subtype generalization failure to identify potential directions for future model development. We selected the best-performing model, ResNet, from previous experiments, trained it on (a) subset, and evaluated it on the (f) subtype for this analysis. The focus of this study was on musical features, including rhythmic patterns, tonal distributions, and low-level audio descriptors such as loudness, which were extracted using the Essentia \cite{bogdanov2013essentia} library.

\begin{figure}[h]
    \centering
    \includegraphics[width=0.85\linewidth]{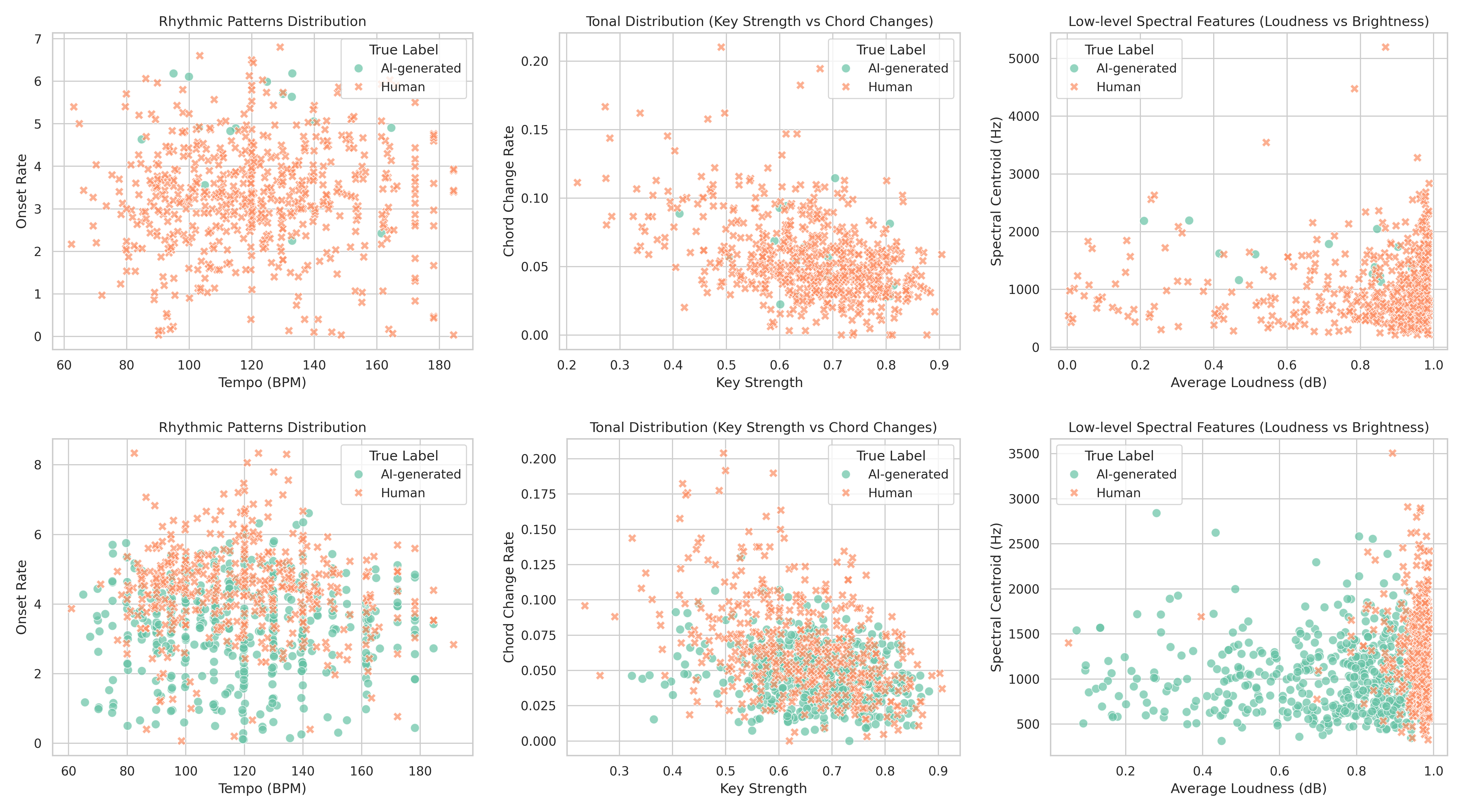}
    \caption{Feature distribution in the (f) subtype comparison between correctly classified and misclassified samples. The top row represents correctly classified samples, while the bottom row represents misclassified samples.}
    \label{fig:ood f}
\end{figure}

To examine the differences between correctly classified and misclassified samples, we analysed representative features from three categories: tempo and onset rate for rhythm-related characteristics; key strength and chord change rate for tonal distribution; and loudness and spectral centroid for low-level physical features, as illustrated in Figure \ref{fig:ood f}.

Our analysis indicates that high-level musical features are not strongly distinguishable, although misclassified samples tend to be human-composed music incorrectly classified as AI-generated more often than vice versa. At the feature level, this suggests that in-domain training did not sufficiently capture less obvious high-level characteristics.

\begin{figure}[h!]
    \centering
    \includegraphics[width=0.7\linewidth]{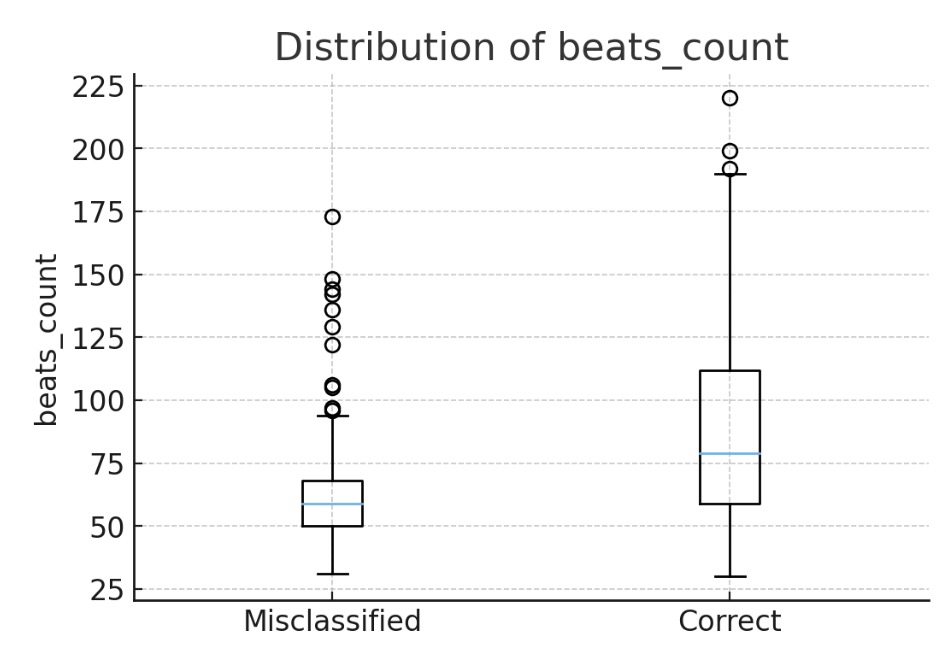}
    \caption{Beats count difference between correctly classified and misclassified samples.}
    \label{fig:beat_count}
\end{figure}

Consequently, we examined lower-level descriptors and found that beat counts show a notable difference between correctly classified and misclassified samples, as illustrated in Figure \ref{fig:beat_count}. Specifically, misclassified samples exhibit significantly fewer beats with mean as 60.6 compared to correctly classified ones with mean as 88.4, indicating that the model learns beats more successfully. This difference is statistically highly significant with p-value lower than 1e-50. This finding is consistent with a musical perspective, as the subsets comprise different instruments that are inherently distinguished by their rhythmic characteristics. Therefore, future models should focus on effectively capturing a broader range of musical features during in-domain training to improve generalisation.

\subsubsection{Preliminary explainable AI techniques}
To conduct a preliminary investigation into why certain methods fail to correctly predict specific cases, we employ the Grad-CAM explainable AI technique \cite{Selvaraju_2019}. For this analysis, we utilise the ResNet on our M6 dataset, aiming to provide initial insights into potential directions for improving MGMD model development.

\begin{figure}[h!]
    \centering
    \includegraphics[width=0.85\linewidth]{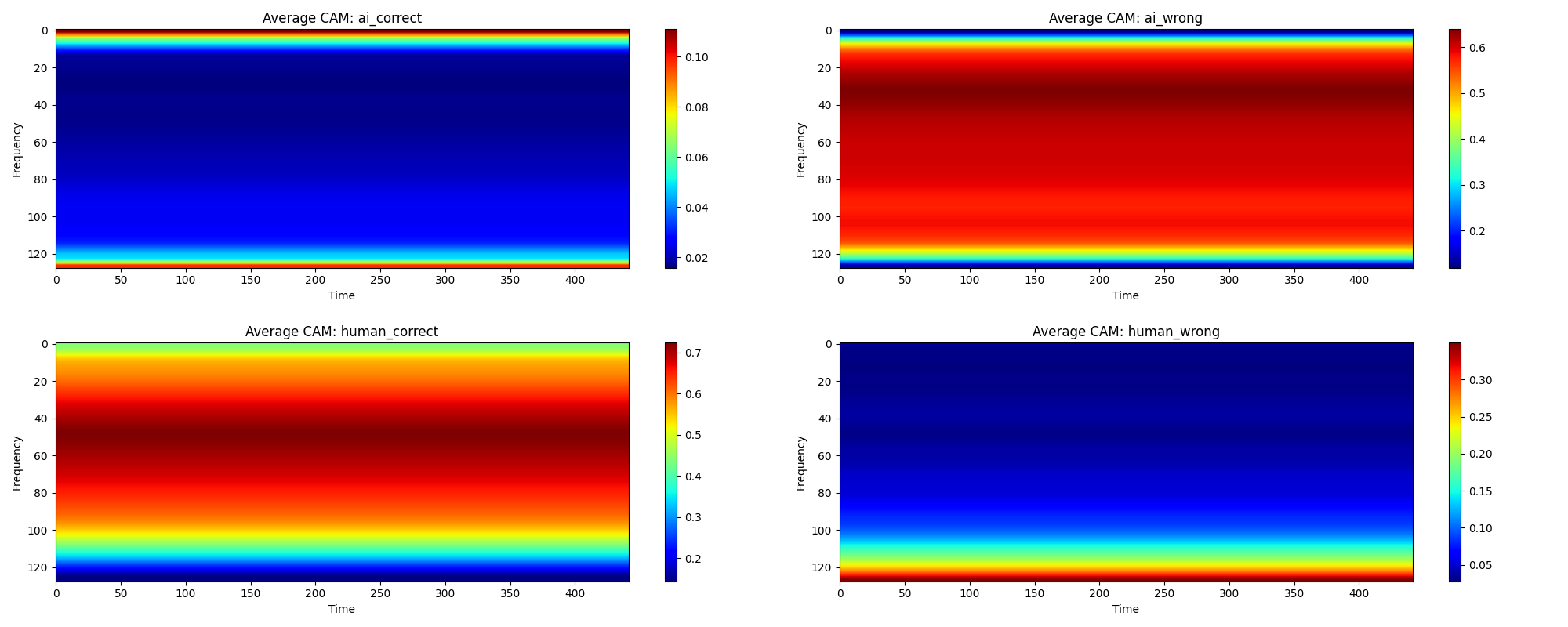}
    \caption{Average Class Activation Maps (CAMs) for AI and human decisions across correct and incorrect conditions for ResNet trained and tested on (a) subtype of M6. The top row shows AI samples (left: correctly classified; right: misclassified), and the bottom row shows human samples (left: correctly classified; right: misclassified). Warmer colors indicate stronger activations.}
    \label{fig:cam}
\end{figure}

Figure \ref{fig:cam} presents the average class activation maps (CAMs) for correctly classified MG and human-produced samples, derived from a ResNet model trained and tested on a subset of the dataset. The results indicate that extreme high and low frequency activations are more prominent in AI-related CAMs while human-related CAMs display stronger activations in the mid-frequency range. This pattern suggests that the model differentiates MG from human-created content primarily based on frequency-related features with AI-associated decisions rely more heavily on high-frequency, while human-associated decisions emphasize mid-frequency. It is important to note, however, that these CAM patterns reflect the model’s internal perception of discriminative features rather than the true frequency composition of the samples. This does not imply that AI-generated samples objectively contain more high-frequency content. Rather, the model may have learned to overemphasize such high-frequency cues as a heuristic for identifying AI-generated data. From Table \ref{tab:music_type_features} and \ref{tab:music_type_high_features}, there are still distinguishable features that model could learn for this task.

Therefore, this analysis should be regarded as a preliminary, high-level observation suggesting that additional, more meaningful musical features—beyond superficial spectral characteristics—should be considered in MGM detection. Future work could apply more specialized XAI techniques to examine the higher-level musical representations learned by such models.

\subsection{Influence of machine generators}
Another widely discussed approach in the detection of machine-generated content is watermarking in specific generators \cite{ren2024sok}. It has been argued that generators exhibit characteristic output distributions, leading to recurring features across their samples—sometimes referred to as authorship—which can introduce bias into the detection process \cite{korabandi2025aigc}. In the M6 dataset, multiple generators are included. Taking AMG, MG, and MGPT as examples, we evaluate both ResNet and ViT models on type (a) and (d), trained on the same data distribution, with results reported in Table~\ref{tab:model_generator}. The findings show that detection accuracy varies consistently across generators, with AMG being the easiest to detect and MGPT the most challenging, while human audio falls in between. This stable ranking across architectures suggests that each generator leaves identifiable acoustic fingerprints that detectors exploit, with MGPT producing samples more similar to human audio and thus reducing detection performance. Such generator-specific fingerprints pose an additional challenge, as detectors may overfit to these watermark-like artifacts, hindering their generalisation capability.

\begin{table}[h]
\centering
\begin{tabular}{l c}
\hline
\textbf{Model–Generator} & \textbf{Accuracy} \\
\hline
ResNet–AMG   & 0.995 \\
ResNet–MG    & 0.962 \\
ResNet–MGPT  & 0.924 \\
ResNet–human & 0.989 \\
ViT–AMG      & 1.000 \\
ViT–MG       & 0.991 \\
ViT–MGPT     & 0.970 \\
ViT–human    & 0.932 \\
\hline
\end{tabular}
\caption{Model performance across different generators.}
\label{tab:model_generator}
\end{table}

\subsection{Future research direction}
As a foundational dataset for AI-generated music detection, this dataset encourages further research at the intersection of AI and music, including fields such as music feature extraction analysis and the development of AI-generated tools. The varied sources within the M6 dataset are expected to enhance the robustness of future detectors and provide new perspectives for AI generation tools. Beyond AI-generated music detection, the M6 dataset comprises different music sources, each subtype of which can be utilised for various tasks. For instance, subtype (a) can be employed for instrumental recognition \cite{poudel2021classification}, especially with additional synthesized data that introduces a different distribution for the classification task. Subtypes (b) and (c) include multilingual lyrics, offering resources for the lyric-melody consistency task, which is crucial for generalising tools from lyric-to-melody models \cite{zhang2022relyme}, and vice versa \cite{tian2023unsupervised}. Subtype (d) encompasses multiple music types, contributing to the publicly important task of style recognition \cite{fu2010survey}, with synthesized components. Subtype (e) echos with challenges related to generating long-sequence data with consistent music style. Additionally, the presence of multiple AI sources within the dataset can aid in recognising generative attribution, a well-known task in audio deepfake detection \cite{bontcheva2024generative}. Moreover, since data imbalance is a common issue in deepfake detection, data stratification strategies could be employed in future model training. For example, upsampling or downsampling can be applied to different data categories, or tailored loss functions can be designed to enable more heuristic sampling of minority cases \cite{tan2025adaptive}.  While AI-generated music detection is a primary focus of our dataset, its applications are not limited to this field alone.

\section{Conclusion}
\label{sec:conclusion}

We have introduced the M6 database for the MGMD task, providing a detailed analysis of this database and comparing it with existing datasets, highlighting its comprehensive sources as well as its broad coverage. Additionally, we have established a baseline performance using common basic models. We question whether transformer-based models are overly sophisticated for this task due to their emphasis on contextual modelling and advocate for further discussion on the most suitable feature extraction methods required. 

Building on these insights, we identify several challenges that should be addressed in future work. We advocate for the development of more robust models that can improve out-of-domain performance. Furthermore, we call for greater model explainability, moving beyond simple binary classification outputs to provide more meaningful interpretations of model decisions. Looking ahead, multimodal models that incorporate lyrics and hierarchical models for handling long-term music structure also require great attention.

\bibliography{sample}
\section{Dataset availability}
See dataset here: \url{https://huggingface.co/datasets/yl7622/M6}
\section{Funding}
This research was partially supported and funded by the Munich Center for Machine Learning and the Munich Data Science Institute.
\section{Author contributions statement}

Yupei Li: lead the project, write manuscripts, do experiments;
Hanqian Li: do experiments;
Lucia Specia: supervision, check manuscripts, provide GPUs;
Bjorn Schuller: main supervision, re-write manuscripts, funding

\end{document}